\documentclass[twocolumn,prl,aps,showpacs]{revtex4}
\usepackage{amsmath}
\usepackage{amsfonts}
\usepackage{graphicx}
\usepackage{textcomp}
\usepackage[colorlinks=true,linkcolor=red,anchorcolor=red,citecolor=blue,urlcolor=black]{hyperref}

\begin{document}

\title{Extrinsic anomalous Hall conductivity of a topologically nontrivial
conduction band}

\author{Hai-Zhou Lu and Shun-Qing Shen}

\affiliation{Department of Physics and Centre of Theoretical and Computational Physics, The University of Hong Kong,
Pokfulam Road, Hong Kong, China}

\begin{abstract}
A key step towards dissipationless transport devices is the quantum anomalous
Hall effect, which is characterized by an integer quantized Hall conductance
in a ferromagnetic insulator with strong spin-orbit coupling. In this work,
the anomalous Hall effect due to the impurity scattering, namely the
extrinsic anomalous Hall effect, is studied when the Fermi energy
crosses with the topologically nontrivial conduction band of a quantum
anomalous Hall system. Two major extrinsic contributions, the side-jump
and skew-scattering Hall conductivities, are calculated using the
diagrammatic techniques in which both nonmagnetic and magnetic scattering
are taken into account simultaneously. The side-jump Hall conductivity
changes its sign at a critical sheet carrier density for the nontrivial phase, while it remains sign unchanged for the trivial
phase. The critical sheet carrier densities estimated with realistic
parameters lie in an experimentally accessible range. The results
imply that a quantum anomalous Hall system could be identified
in the good-metal regime.
\end{abstract}


\date{\today }

\maketitle



\begin{figure}[htbp]
\centering \includegraphics[width=0.3\textwidth]{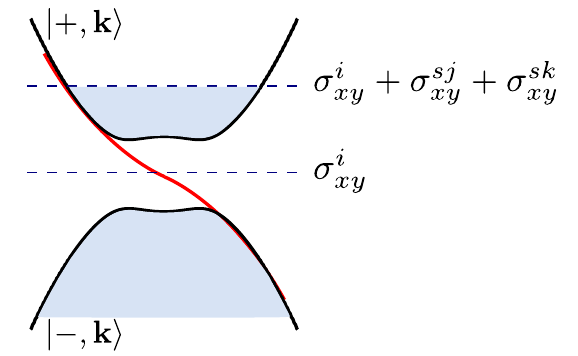} \caption{$|\pm,\mathbf{k}\rangle$ represent conduction and valence bands of
a quantum anomalous Hall system, respectively. When the Fermi energy
(dashed) lies between the two bands, the
anomalous Hall conductivity is governed dominantly by the intrinsic mechanism
($\sigma_{xy}^{i}$), which is due to the chiral edge states (red
solid). When the Fermi energy crosses with the conduction (or valence)
band, the extrinsic mechanisms ($\sigma_{xy}^{sj}$ for side-jump
and $\sigma_{xy}^{sk}$ for skew-scattering) also contribute to the anomalous
Hall conductivity.}
\label{fig:model}
\end{figure}

\emph{Introduction} - The anomalous Hall effect appears in ferromagnets
as a transverse current induced by a longitudinal electric field.
Different from the ordinary Hall effect, it is not driven by the Lorentz
force acting on charge carriers in a magnetic field. Instead, it stems from
the interplay of the spin-orbit coupling and time-reversal symmetry
breaking \cite{Nagaosa10rmp}. The anomalous Hall conductance has
two distinct contributions, from the extrinsic and intrinsic mechanisms.
The extrinsic mechanism originates from electrons near the Fermi surface when they are scattered by impurities. The intrinsic mechanism, on the
contrary, is given by the Berry curvature of electrons below
the Fermi surface, as a consequence of the spin-orbit coupling induced
topological properties in Bloch bands \cite{Xiao11rmp}. In particular,
the intrinsic anomalous Hall conductance could be quantized in units
of the conductance quantum when the Fermi surface lies in the gap
between energy bands. Known as the quantum anomalous Hall effect\cite{Liu08prl,Yu10sci},
the phenomenon is a key step towards dissipationless quantum transport without magnetic field, and
thus has attracted much efforts for its experimental realization \cite{Chang13am,Checkelsky12natphys}.
One promising proposal is to magnetically dope a quantum spin Hall
system \cite{Kane05qsh,Bernevig06sci}, which can be regarded as two
time-reversed copies of the quantum anomalous Hall system. The two copies have exactly opposite
Hall conductances that cancel with each other. The magnetic doping
\cite{Hor10prb,Chen10sci,Wray11natphys}, which breaks time-reversal
symmetry, can lift the cancellation and give rise to the quantum anomalous
Hall effect. However, the doping and inheriting defects always bring
extra carriers, which shift the Fermi energy out of the gap and into
an energy band where electron transport suffers from impurity scattering.
In this situation, the extrinsic mechanisms also becomes relevant
(see Fig. \ref{fig:model}), but was never addressed.

In this work, we study the extrinsic anomalous Hall effect of the
conduction band of a quantum anomalous Hall system. With the help
of the Kubo formula and Feynman diagrams, we calculate the side-jump
and skew-scattering contributions to the Hall conductivity, two major
extrinsic mechanisms. We find that the side-jump Hall conductivity
could change sign at an experimentally accessible sheet carrier density in the topologically nontrivial phase, while its sign remains unchanged in the trivial case. The skew-scattering Hall conductivity show a similar behavior when nonmagnetic scattering dominates. The sign-changing feature may serve as a signature for the quantum anomalous Hall system in a dirty device.

\emph{Model} - The minimal model for the quantum anomalous Hall system
takes the form
\begin{eqnarray}
H=\gamma(k_{x}\sigma_{x}+k_{y}\sigma_{y})+(\frac{\Delta}{2}-Bk^{2})\sigma_{z},\label{Hamiltonian}
\end{eqnarray}
where $\gamma=v\hbar$, $v$ is the effective velocity, $(k_{x},k_{y})$
is the wave vector, and $k^{2}=k_{x}^{2}+k_{y}^{2}$. $\sigma_{x/y}$
are Pauli matrices. $\Delta=m_{0}\pm gM$, where $m_{0}$ is a gap opened
at $k=0$ for the hybridized surface states of a three-dimensional
topological insulator \cite{Lu10prb}. $gM$ describes the magnetic
doping-induced exchange field, which may effectively modulate $\Delta$
between positive and negative values for a given $m_{0}$ \cite{Liu08prl,Yu10sci}.
The $Bk^{2}\sigma_{z}$ term is a topological correction \cite{Shen11spin,Shen-book}
to the usual minimal model for the anomalous Hall effect \cite{Onoda06prl,Sinitsyn07prb,Yang11prb,Culcer11prb,Nomura11prl,Chu11prb}.
$H$ has one conduction band $|+,\mathbf{k}\rangle$ and one valence
band $|-,\mathbf{k}\rangle$, with the dispersion $\epsilon_{\pm\bold{k}}=\pm\sqrt{(\Delta/2-Bk^{2})^{2}+\gamma^{2}k^{2}}$.
We assume that the Fermi energy $E_{F}$ always crosses with the conduction
band, where $E_{F}$ is measured from the Dirac point at $\epsilon_{\pm\mathbf{k}}=0$.

\emph{Intrinsic Hall conductance} - When the Fermi energy lies in
the gap, the longitudinal conductance is zero, i.e., the system is
insulating as the valence band is fully filled. However, there may exist an intrinsic quantum anomalous Hall conductance \cite{Lu10prb,supp}
\begin{eqnarray}
\sigma_{xy}^{i(-)}=-\frac{e^{2}}{2h}[\mathrm{sgn}(\Delta)+\mathrm{sgn}(B)],\label{Chern-}
\end{eqnarray}
where $\mathrm{sgn}(x)=1$ if $x>0$, $\mathrm{sgn}(x)=-1$ if $x<0$,
and $\mathrm{sgn}(x)=0$ if $x=0$. If $\Delta B>0$, the model is
in the nontrivial phase as $\sigma_{xy}^{i(-)}$ has an
integer anomalous Hall conductance in units of $e^{2}/h$.
The integer, which corresponds to the chiral edge states in the gap, is the Chern number from the nontrivial topological properties of the valence band. The conduction band always has an opposite Chern number compared to the valence band and thus also carries the nontrivial topological properties.
If $\Delta B<0$, the model
is called topologically trivial with a zero anomalous Hall conductance.
The importance of $Bk^{2}\sigma_{z}$ term deserves to be emphasized:
With it, the solution of the in-gap chiral edge states can be found explicitly at an open edge \cite{Zhou08prl}.
Without the term, Eq. (\ref{Chern-}) gives a half-integer
anomalous Hall conductance \cite{Qi06prb,Onoda06prl,Chu11prb}, and there is
no edge-state solution at an open edge although the Jackiw-Rebbi bound
state is allowed near a domain wall at which $\Delta$ changes sign
\cite{Jackiw76prd}.

\emph{Side-jump Hall conductivity} - Breaking time-reversal symmetry is indispensable for the anomalous Hall effect, so the anomalous Hall conductivity must depend on the time-reversal breaking terms in the model, such as $(\Delta/2-Bk^{2})\sigma_{z}$ and magnetic scattering. The side-jump mechanism is related to the impurity scattering but not proportional to the total impurity
concentration and scattering strength \cite{Nagaosa10rmp}. As we will
see, the side-jump extrinsic Hall conductivity is proportional to
\begin{eqnarray}\label{costheta}
\cos\theta_F=\frac{\Delta/2-Bk_{F}^{2}}{E_{F}}
\end{eqnarray}
where $k_{F}$ is the Fermi wave vector. This means that the side-jump
Hall conductivity could change sign at a critical Fermi wave vector
$k_{C}=\sqrt{\Delta/2B}$ if $\Delta B>0$, that is, if the system
is in the nontrivial phase. In contrast, the Hall conductivity
is monotonic if $\Delta B<0$, i.e., if the system is trivial.
With
the critical $k_{C}$, we can find the critical sheet carrier density
$n_{C}=k_{C}^{2}/(4\pi)$ and Fermi energy $E_C=v\hbar \sqrt{\Delta/2B}$.
We estimate $k_C$, $n_C$, and $E_C$ with the experimental
fitting data for topological insulator thin films and calculated parameters
for HgTe quantum well, which are proposed host materials for the quantum
anomalous Hall system \cite{Liu08prl,Yu10sci}.
Table \ref{tab:kc} shows the critical values of $k_{C}$ and $n_{C}$.
In 10 nm $n$-type Bi$_{2}$Se$_{3}$ thin films \cite{Kim12natphys},
only the surface states
are populated for sheet carrier density below $5\sim7.7\times10^{12}$/cm$^{2}$.
Thus, most critical $n_{c}$ in Table. \ref{tab:kc} lie inside an
experimentally accessible regime.
The critical $n_C$ for HgTe is
even much smaller. The above discussion shows again that the $Bk^{2}\sigma_{z}$
term cannot be underestimated in a quantitative analysis.
The first-principles calculations show that the bulk band minima will be pushed to higher energies in Bi$_2$Se$_3$ thin films \cite{Yu10sci} (e.g., about 0.3 eV for 5 QL and 0.4 eV for 3 QL), higher than corresponding $E_C$ in Tab. \ref{tab:kc}.
The higher-order terms other than those in $H$ may shift the critical points, but will not qualitatively affect the sign changing feature as long as they preserve time-reversal symmetry. Also, the Berry phase is related to Eq. (\ref{costheta}) by $\pi(1\pm \cos\theta_F)$. It is known that the $\pi$
Berry phase leads to weak antilocalization in the longitudinal transport \cite{Suzuura02prl}. The vanishing of $\cos\theta_F$ at the critical carrier density was predicted to give weak antilocalization if $\Delta B>0$ \cite{Lu11prb}.

\begin{table}[htbp]
\caption{Calculated critical Fermi wave vector $k_{C}$, sheet carrier density
$n_{C}$, and Fermi energy $E_C$ with $\Delta$, $B$, and $v_F$ from experimental fitting data and
$k\cdot p$ calculations. Entries 1-4 are by Zhang \emph{et al}.\cite{Zhang10natphys},
Entry 5 is by Sakamoto \emph{et al}.\cite{Sakamoto10prb}, Entry 6
is by Konig \emph{et al}.\cite{Konig08jpsj}. QL for quintuple layer
is about 1 nm. $\Delta$ in eV, $B$ in eV\AA$^{2}$, $v$ in 10$^5$m/s, $k_{C}$ in
\AA$^{-1}$, $n_{C}$ in 10$^{12}$/cm$^{2}$, and $E_C$ in eV. The parameters
in magnetically-doped samples may be different.
In magnetically-doped case, the exchange field could reduce $\Delta$, leading to smaller $n_C$.}
\label{tab:kc}%
\begin{ruledtabular}
\begin{tabular}{ccccccc}
  Sample & $\Delta $  &  $B$ & $v$ & $k_C$ & $n_C$& $E_C$  \\
\hline
  2QL Bi$_2$Se$_3$ & 0.252 & 21.8 & 4.71 & 0.076  & 4.6& 0.24 \\
  3QL Bi$_2$Se$_3$ & 0.138 & 18 & 4.81  & 0.062 & 3.1& 0.20\\
  4QL Bi$_2$Se$_3$ & 0.07 & 10 & 4.48  & 0.059 & 2.8& 0.17 \\
  5QL Bi$_2$Se$_3$ & 0.041 & 5.0 & 4.53  &0.064 & 3.3& 0.19 \\
  3QL Bi$_2$Se$_3$ & 0.34 & 18 & 4.4  & 0.1  & 7.5& 0.28\\
  7 nm HgTe & -0.01 & -68.6 & 5.54  & 0.009 & 0.058& 0.03 \\
\end{tabular}
\end{ruledtabular}
\end{table}

\begin{figure}[htbp]
\centering \includegraphics[width=0.23\textwidth]{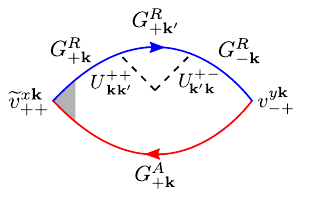} \centering
\includegraphics[width=0.23\textwidth]{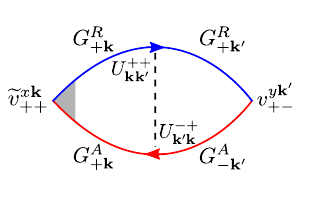} \centering \includegraphics[width=0.23\textwidth]{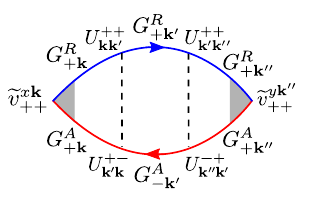}
\centering \includegraphics[width=0.23\textwidth]{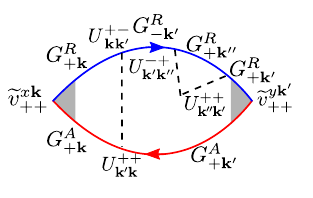} \centering
\includegraphics[width=0.23\textwidth]{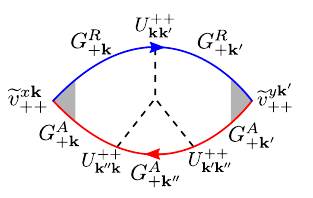} \caption{The diagrams for the extrinsic Hall conductivity can be summarized to only
5 of them. Top four: Side-jump contribution. Bottom: Skew-scattering
contribution. $\pm$ are band indices. $\mathbf{k},\mathbf{k}',\mathbf{k}''$
are wave vectors. The left-bound and right-bound arrowed lines stand for
retarded ($G^{R}$) and advanced ($G^{A}$) Green's functions, respectively.
$U_{\mathbf{k},\mathbf{k}'}$ is scattering matrix elements. The dashed
lines represent the correlation between scattering. $v^{x/y}$ are
the bare velocities. The shadow area and tilde represent the vortex
correction to the bare velocities. }
\label{fig:side-jump}
\end{figure}

The extrinsic Hall conductivity can be calculated by the Kubo formula
in terms of the Feynman diagrams, where the impurity scattering is
treated as the perturbation to the states $|\pm,\mathbf{k}\rangle$.
The diagrams for the extrinsic anomalous Hall conductivity (see Fig.
\ref{fig:side-jump}) have been systematically developed \cite{Sinitsyn07prb} and applied to the model without the $Bk^{2}\sigma_{z}$
term \cite{Sinitsyn07prb,Yang11prb}. Here we generalize
the diagrammatic calculation by including the extra $Bk^{2}\sigma$ term, and considering the nonmagnetic
and magnetic impurities simultaneously. Despite lengthy calculation,
we arrive at a compact form for the side-jump Hall conductivity \cite{supp}
\begin{eqnarray}
\sigma_{xy}^{sj}=-\frac{e^{2}}{h}\cos\theta_{F}\left[\frac{2\alpha}{1-\alpha}+\frac{3\alpha^{2}\eta_{B}}{2(1-\alpha)^{2}}\right],\label{sigma-sj-24}
\end{eqnarray}
where $\cos\theta\equiv(\Delta/2-Bk_{F}^{2})/E_{F}$,
\begin{eqnarray}
\alpha=\frac{\frac{1}{2}(1-V_{m}/V_0)\sin^{2}\theta_{F}}{2-\sin^{2}\theta_{F}+(V_{m}/V_0)(2+\sin^{2}\theta_{F})},\label{alpha-beta}
\end{eqnarray}
and $\eta_{B}=1-2Bk_{F}/(\gamma\tan\theta_{F})$. $V_{0}\equiv n_{0}u_{0}^{2}$
and $V_{m}\equiv n_{m}u_{m}^{2}$ here are of physical meanings \cite{Lu11prl,Shan12prb}.
$n_{0}$ and $n_{m}$ are the concentrations of nonmagnetic
and magnetic impurities, respectively. $u_{0}$ and $u_{m}$ are spatially-averaged strengths for the nonmagnetic and each component of the magnetic scattering, respectively. We have assumed isotropic
magnetic scattering.

\begin{figure}[htbp]
\centering \includegraphics[width=0.48\textwidth]{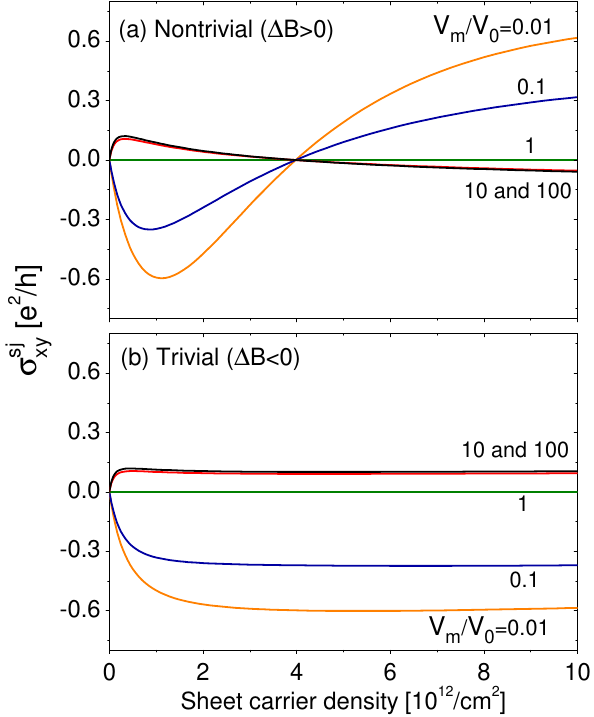}
\caption{The side-jump Hall conductivity as a function of sheet carrier density.
For all cases $\Delta=0.1$ eV and $v=4\times10^{5}$ m/s. $B=10$
eV\AA$^{2}$ for the nontrivial (quantum anomalous Hall) case (a) and $B=-10$
eV\AA$^{2}$ for the trivial case. For better comparison, the parameters
are not adopted directly from those in the experiments but of the
same orders. $n_C$ is about $4\times 10^{12}$/cm$^2$ in this case. }
\label{fig:sigma-sj-24}
\end{figure}

Figure \ref{fig:sigma-sj-24} shows $\sigma_{xy}^{sj}$ in Eq. (\ref{sigma-sj-24})
for the quantum anomalous Hall and trivial cases. The horizontal axis is the sheet carrier
density, which can be determined by the ordinary Hall measurement.
As expected, $\sigma_{xy}^{sj}$ changes sign at the critical value
of $n_C$ for the quantum anomalous Hall case [Fig. \ref{fig:sigma-sj-24}
(a) $\Delta B>0$], while its sign remains unchanged for the trivial
[Fig. \ref{fig:sigma-sj-24} (b) $\Delta B<0$] case. Besides,
Fig. \ref{fig:sigma-sj-24} also shows an impurity-related sign change
in $\sigma_{xy}^{sj}$ as the ratio $V_{m}/V_{0}$ changes. $V_{0}$ and $V_m$ measure the nonmagnetic and each component of the magnetic scattering, respectively.
Although the side-jump Hall conductivity does not depends on the total
scattering strength and impurity concentration, it may depend on the
relative weight of different types of scattering. Varying $V_{m}/V_{0}$
gives rise to a sign change in $\sigma_{xy}^{sj}$ in both trivial
and nontrivial cases in Fig. \ref{fig:sigma-sj-24}. This impurity-dependent
sign change happens exactly at $V_{m}=V_{0}$, at which both $\alpha$ in Eq. (\ref{alpha-beta}) and $\sigma_{xy}^{sj}$ vanish.

\emph{Skew-scattering Hall conductivity} - The skew-scattering Hall
mechanism originates from the asymmetric scattering induced by the
spin-orbit coupling. The leading order of the skew-scattering Hall
conductivity can be calculated from the diagram in Fig. \ref{fig:side-jump},
and found as \cite{supp}
\begin{eqnarray}\label{sigma-sk}
\sigma_{xy}^{sk}=-\frac{e^{2}}{h}\frac{E_{F}(\eta_{B}\sin^{2}\theta_{F})^{2}(V_{3}^{0}\cos\theta_{F}-V_{3}^{z})}{(1-\alpha)^{2}[V_{0}(2-\sin^{2}\theta_{F})+V_{m}(2+\sin^{2}\theta_{F})]^{2}},\nonumber
\end{eqnarray}
where the third-order impurity scattering correlations are defined
as $V_{3}^{0}\equiv\langle U_{\mathbf{k}\mathbf{k}'}^{0}U_{\mathbf{k}'\mathbf{k}''}^{0}U_{\mathbf{k}''\mathbf{k}}^{0}\rangle$,
$V_{3}^{z}\equiv\langle U_{\mathbf{k}\mathbf{k}'}^{z}U_{\mathbf{k}'\mathbf{k}''}^{z}U_{\mathbf{k}''\mathbf{k}}^{z}\rangle$,
with $0$ and $z$ corresponding to the nonmagnetic elastic scattering
and $z$ component of the magnetic scattering. $x$ and $y$ components
are abandoned in the presence of the in-plane rotational symmetry.
The above equation shows that only the nonmagnetic scattering
part with $V_{3}^{0}$ is proportional to $\cos\theta_{F}$, so the
skew-scattering shows the similar sign-changing feature only in absence
of the magnetic scattering. $V_{3}^{0/z}$ correspond to the correlation
of three scattering events by one single impurity, so $V_{3}^{0/z}$
are linearly proportional to the impurity concentration $n_{0/m}$.
Meanwhile the second-order scattering $V_{0/m}$ are
also linearly proportional to $n_{0/m}$, so roughly speaking $\sigma_{xy}^{sk}$
is inversely proportional to the impurity concentration. For this
reason, the skew-scattering Hall conductivity $\sigma_{xy}^{sk}$
is suppressed in a dirty metal.

\emph{Experimental implication} - Because only the side-jump Hall
conductivity always shows the sign change in the nontrivial
phase, it is necessary to extract it among the three major contributions
to the anomalous Hall conductivity. In principle, the intrinsic, side-jump,
and skew-scattering mechanisms can be distinguished in experiments
\cite{Nagaosa10rmp}. Empirically, the skew-scattering mechanism dominates
in the high-conductivity regime where the longitudinal conductivity
$\sigma_{xx}>10^{6}$ ($\Omega$ cm)$^{-1}$ and the anomalous Hall
resistivity is proportional to the longitudinal resistivity \cite{Nagaosa10rmp}.
Considering the low conductivity in all the samples of Bi$_{2}$Se$_{3}$
family, the skew-scattering mechanism looks quite irrelevant for the
recent experiments \cite{Chang13am}. In the good-metal regime where
the Hall conductivity is independent of the longitudinal conductivity
and $10^{4}<\sigma_{xx}<10^{6}$ ($\Omega$ cm)$^{-1}$, both the
intrinsic and side-jump mechanisms could contribute. When the Fermi
surface intersects the conduction band, the intrinsic mechanism also
contribute a Hall conductivity \cite{Lu10prb,supp}
\begin{eqnarray}
\sigma_{xy}^{i}=\sigma_{xy}^{i(-)}+\left.\sigma_{xy}^{i(+)}\right|_{E_{F}}=-\frac{e^{2}}{2h}[\mathrm{sgn}(B)+\cos\theta_{F}].\label{sigma-int-cv}
\end{eqnarray}

\begin{figure}[htbp]
\centering \includegraphics[width=0.48\textwidth]{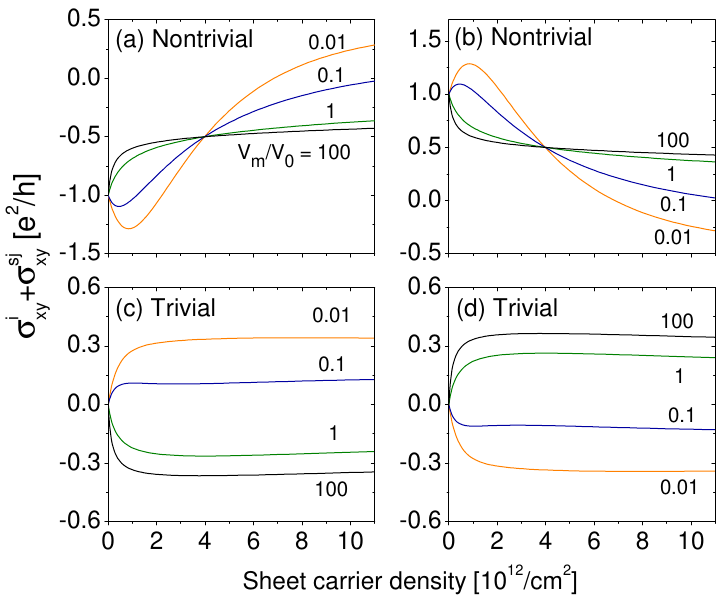}
\caption{The sum of intrinsic and side-jump Hall conductivities as functions
of sheet carrier density for different $V_{m}/V_{0}$. A square device is assumed so conductivity is equivalent to conductance. Parameters: $v_{F}=4\times10^{5}$ m/s, (a) $\Delta=0.1$
eV, $B=10$ eV.\AA$^{2}$; (b) $\Delta=-0.1$ eV, $B=-10$ eV.\AA$^{2}$;
(c) $\Delta=-0.1$ eV, $B=10$ eV.\AA$^{2}$; (d) $\Delta=0.1$
eV, $B=-10$ eV.\AA$^{2}$. }

\label{fig:sigma-sj-approx-int}
\end{figure}

Figure \ref{fig:sigma-sj-approx-int} shows the sum of $\sigma_{xy}^{i}$
and $\sigma_{xy}^{sj}$ as functions of the sheet carrier density
of the conduction band. For the nontrivial phase, the
curves of different $V_{m}/V_{0}$ cross at the critical sheet carrier
density where $\sigma_{xy}^{sj}$ changes sign and the value of the
Hall conductivity is shifted to
$ \left.\sigma_{xy}^{i}\right|_{n_{C}}+\left.\sigma_{xy}^{sj}\right|_{n_{C}}=-\frac{e^{2}}{2h}\mathrm{sgn}(B)$. This shifting can be canceled if the time-reversed partner of $H$ is also considered, which gives $\mathrm{sgn}(-B)$. In contrast, those for the trivial case do not cross [Figs. \ref{fig:sigma-sj-approx-int} (c) and (d)]. This crossing
could provide an extra signature of the nontrivial phase, if the
relative strength of the nonmagnetic and magnetic doping fluctuates accidentally or can be tuned in a controlled way from sample to sample. Also,
the intrinsic and side-jump contributions can be separated by defining
the intrinsic contribution as the extrapolation of the ac Hall conductivity
to zero frequency in the limit of $\tau_{tr}\rightarrow\infty$, with $1/\tau_{tr}\rightarrow0$
faster than $\omega\rightarrow0$ \cite{Nagaosa10rmp}. $\tau_{tr}$ is
the transport time, which can be extracted from the longitudinal
conductivity.

\emph{Summary} - We show that extrinsic anomalous Hall conductivity in a topologically nontrivial conduction band (i.e., in the quantum anomalous Hall phase) exhibits different behaviors from those in a trivial band. More specifically, the side-jump extrinsic Hall conductivity changes sign at a critical sheet carrier density only in the nontrivial phase. When varying the ratio between nonmagnetic and magnetic scattering, the side-jump Hall conductivities cross at the critical sheet carrier density.
The skew-scattering Hall conductivity shows similar
behaviors when the nonmagnetic scattering overwhelms the magnetic scattering.
These features may help future experiments that explore the quantum anomalous Hall systems.

This work was supported by the Research Grant Council of Hong Kong
under Grant No. HKU705111P.


\end{document}